\documentclass{aa}
\usepackage{graphicx}
\usepackage{txfonts}
\usepackage{natbib}
\bibpunct{(}{)}{;}{a}{}{,} 
\begin{document}
%
%
 \newcommand{\betacrbbold}{\mbox{\boldmath $\beta$}~CrB} 

 \newcommand{\betacrb}{$\beta$~CrB}
 \newcommand{\tenaql}{10~Aql}
 \newcommand{\gammaequ}{$\gamma$~Equ}

 \newcommand{\thetald}{$\theta_{\rm LD}$}
 \newcommand{\fbol}{$f_{\rm bol}$}
 \newcommand{\acir}{$\alpha$~Cir}
 \newcommand{\ie}{i.e.}
 \newcommand{\eg}{e.g.}
 \newcommand{\cf}{cf.}
 \newcommand{\kms}{km\,s$^{-1}$}
 \newcommand{\teff}{$T_{\rm eff}$}
 \newcommand{\logg}{$\log g$}
 \newcommand{\feh}{[Fe/H]}
 \newcommand{\msun}{${\rm M}_{\odot}$}
 \newcommand{\percent}{\,{\%}}

%


%
%
\title{The radius and effective temperature of the binary Ap~star $\beta$~CrB from CHARA/FLUOR and VLT/NACO observations\thanks{Based on observations made with ESO telescopes at the La Silla Paranal Observatory,
under ESO DDT program 281.D-5020(A).}}
\titlerunning{FLUOR and NACO observations of \betacrb~A and B}
\authorrunning{H. Bruntt et al.}
\author{
H.~Bruntt\inst{1,2}
\and
P.~Kervella\inst{1}
\and
A.~M\'erand\inst{3}
\and
I.\ M.\ Brand{\~a}o\inst{4,5}
\and
T.~R.~Bedding\inst{2}
\and
T.~A.~ten Brummelaar\inst{6}  
\and
V.~Coud\'e~du~Foresto\inst{1}  
\and
M.~S.~Cunha\inst{4}
\and
C.~Farrington\inst{6}  
\and
P.~J.~Goldfinger\inst{6}  
\and
L.~L.~Kiss\inst{2,7}
\and
H.~A.~McAlister\inst{6}  
\and
S.~T.~Ridgway\inst{8}  
\and
J.~Sturmann\inst{6}   
\and
L.~Sturmann\inst{6}  
\and
N.~Turner\inst{6}  
\and
P.~G.~Tuthill\inst{2}  
}
\offprints{H. Bruntt}
\mail{bruntt@phys.au.dk}
\institute{
LESIA, CNRS UMR 8109, Observatoire de Paris-Meudon, 5 place Jules Janssen, F-92195 Meudon Cedex, France
\and
Sydney Institute for Astronomy, School of Physics, The University of Sydney, NSW 2006, Australia
\and
European Southern Observatory, Alonso de C\'ordova 3107, Casilla 19001, Santiago 19, Chile
\and
Universidade do Porto, Centro de Astrof\'isica, Rua das Estrelas, PT 4150-762 Porto, Portugal
\and
Departamento de Matem\'atica Aplicada, Faculdade de Ci\^encias, Universidade do Porto, 4169 Porto, Portugal
\and
Center for High Angular Resolution Astronomy, Georgia State University, PO Box 3965, Atlanta, Georgia 30302-3965, USA
\and
Konkoly Observatory of the Hungarian Academy of Sciences, Budapest, Hungary
\and
National Optical Astronomy Observatory, PO box 26732, Tucson, AZ 85726, USA.
}
\date{Received ??-?? 2009 ; Accepted ??-?? ???? }
\abstract
{The prospects for using asteroseismology of rapidly oscillating Ap (roAp) stars are 
hampered by the large uncertainty in fundamental stellar parameters. 
Results in the literature for the effective temperature (\teff) often span a range of 1000~K.}
{Our goal is to reduce systematic errors and improve the \teff\ calibration of Ap stars based 
on new interferometric measurements.}
{We obtained long-baseline interferometric observations of \betacrb\ using the CHARA/FLUOR instrument. 
In order to disentangle the flux contributions of the two components of this binary star, 
we additionally obtained VLT/NACO adaptive optics images.}
{We determined limb darkened angular diameters of $0.699\pm0.017$~mas 
for \betacrb~A (from interferometry) and 
$0.415\pm0.017$~mas for \betacrb~B (from surface brightness-color relations), corresponding to radii of 
$2.63\pm0.09 {\rm R}_\odot$  (3.4\percent\ uncertainty) and 
$1.56\pm0.07\,{\rm R}_\odot$ (4.5\percent). 
The combined bolometric flux of the {A$+$B} components was determined from 
satellite UV data, 
spectrophotometry in the visible and 
broadband data in the infrared. 
The flux from the~B component constitutes $16\pm4$\percent\ of the total flux 
and was determined by fitting an ATLAS9 model atmosphere to the broad-band 
NACO $J$ and $K$ magnitudes. Combining the flux of the~A component with its 
measured angular diameter, we determine the effective temperature 
$T_{\rm eff}\, ({\rm A}) = 7980\pm180$\,K (2.3\percent).}
{Our new interferometric and imaging data enable a nearly model-independent determination 
of the effective temperature of \betacrb~A. 
Including our recent study of \acir, we now have direct \teff\ measurements 
of two of the brightest roAp stars, 
providing a strong benchmark for an improved calibration of the \teff\ scale for Ap stars.  
This will support the use of potentially strong constraints imposed by asteroseismic 
studies of roAp stars.}
%
%
%
%
%
\maketitle
%
%
%
\section{Introduction \label{sec:intro}}

Photometric and spectroscopic determinations of 
the effective temperatures of Ap stars are affected by
systematic errors. This has been corroborated by asteroseismic 
data of rapidly oscillating Ap (roAp) stars in general and, 
more recently, by the first interferometric determination of
the angular diameter of the roAp star \acir\ \citep{bruntt08}.
Unfortunately, the intriguing asteroseismic potential offered 
by roAp stars is strongly compromised by the presence of these 
systematic errors. We therefore seek to make direct measurements
of the radii and effective temperatures of a number of Ap stars
using interferometric and spectrophotometric measurements.
We will first give a brief summary of the properties of
\betacrb\ before describing our observations, data reduction (Sect.~\ref{sec:data})
and analysis (Sect.~\ref{sec:analysis}).

\subsection{A brief history of \betacrb}

\object{$\beta$ CrB} (\object{HD 137909}, \object{HIP 75695})
is one of the brightest, coolest, and best-studied magnetic Ap stars. 
The literature on the star is extensive and we only mention a few of the most
important results here. \betacrb\ has been classified as type A9\,Sr\,Eu\,Cr
star by \cite{renson09}.
Its binary nature was first suggested by \cite{moore1907}, 
and recent determinations of its orbital elements were obtained by 
\cite{tokovinin84} and \cite{north98}. 
From speckle interferometric measurements using narrow-band filters,
\cite{horch04} measured the magnitude difference to be 
2.37\,mag at 551\,nm and 1.99\,mag at 503\,nm.
In the analysis presented in Sect.~\ref{sec:analysis}
we retain the orbital elements obtained by \cite{tokovinin84}, 
as they are in significantly better agreement with our NACO astrometry than \cite{north98}.

\cite{neubauer44} suggested that a third body could be present in the system, 
causing radial velocity variations with a period of $\approx321$\,days, 
but \cite{oetken84}, \cite{kamper90} and \cite{soderhjelm99} excluded this possibility. 
Recently, \cite{muterspaugh06} established an upper limit of 
$\approx10$ to $100$\,$M_J$ (depending on the orbital period) for a possible substellar 
tertiary from differential interferometric astrometry. 
\cite{trilling07} searched for 24 and 70\,$\mu$m infrared excess around 
\betacrb\ using \emph{Spitzer} but did not find any. 
Interestingly, the Spitzer flux they
obtained is significantly below the expected flux at 24\,$\mu$m, and slightly lower (although compatible)
at 70\,$\mu$m. This result could be due to the chosen physical parameters for their stellar atmosphere model.
In the following, we will therefore consider that \betacrb\ is a binary system.

\subsection{Is \betacrb\ a pulsating star?}

Early photometric searches for pulsation in \betacrb\ 
(\eg\ \citealt{weiss89,kreidl91}) gave null results and this
contributed to the discussion of the existence of 
non-oscillating Ap stars (``noAp''; \citealt{kurtz89}).
This has changed since the advent of large telescopes and ultra-stable spectrographs.
Based on spectroscopic time series of a single Fe line, 
\cite{koch02} claimed the first possible detection 
of a pulsation mode in \betacrb\ with a period of 11.5\,min. 
This result was questioned by \cite{hatzes04} and was also not confirmed by \cite{kurtz07}.
However, the good agreement between the independent spectroscopic studies of
\cite{hatzes04}, \cite{kurtz07} and \cite{koch08} confirmed 
that \betacrb\ is indeed an roAp star with a single known low-amplitude mode with period 16.2\,min.
The most robust result was found by \cite{kurtz07}, who used 
2\,hours of high-cadence time-series spectra obtained with
VLT/UVES. They detected a single oscillation frequency
at $1.031$\,mHz ($P=16.2$\,min) with an amplitude of
$23.5\pm2.4$\,\kms\ in the H\,$\alpha$ line and
a higher amplitude in the cesium lines.
Unlike most roAp stars, variation was observed only in 
singly-ionized rare-earth elements, but not doubly ionized lines. 
The abundance analysis done by \cite{kurtz07} on their 
averaged spectrum confirmed earlier investigations by 
\cite{ryab04}. These analyses show that \betacrb\ has an
overabundance of rare-earth elements but only by about 1~dex, 
contrary to the 2--3~dex seen in most roAp stars.

\begin{figure}[t]
\centering
\includegraphics[width=8.7cm, angle=0]{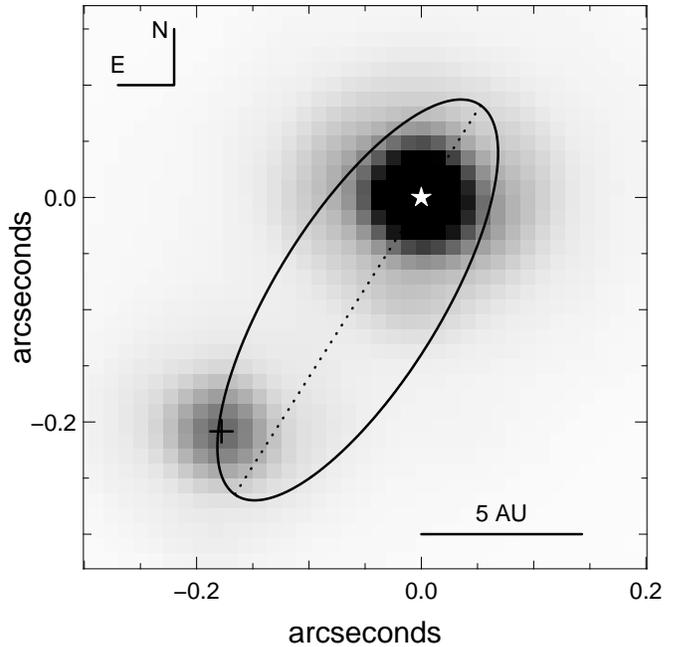}
\caption{Average NACO image of \betacrb~A and B in the $K$ band, together with the binary orbit from \cite{tokovinin84}. The positions of \betacrb~A and B measured on our NACO image are marked with a ``$\star$" and ``+" symbol, respectively. The scale of the two axes are in arc seconds, relative to \betacrb~A. The linear scale in AU is based on the \emph{Hipparcos} parallax.}
\label{betcrbK}
\end{figure}



\section{Observations and data reduction\label{sec:data}}

\subsection{VLT/NACO adaptive optics imaging \label{naco_obs}}

We observed \betacrb\ on 16 June 2008 using the Nasmyth Adaptive Optics System (NAOS; \citealt{rousset03}) of the Very Large Telescope (VLT), coupled to the CONICA infrared camera \citep{lenzen98}, abbreviated as NACO. We selected the smallest available pixel scale of $13.26 \pm 0.03$\,mas/pix \citep{masciadri03}, giving a field of view of 13.6$\arcsec$$\times$13.6$\arcsec$. This small scale resulted in good sampling of the Point Spread Function (PSF).
We employed the $J$ and $K$ filters of NACO, with respective bandpasses of $1.265 \pm 0.25\,\mu$m and $2.18 \pm 0.35\,\mu$m, together with a neutral density filter (labeled ``{\tt ND2\_short}", transmission $\approx 1.5$\percent) to avoid saturation of the detector. The transmission curves of these filters are available on the NACO instrument web page\footnote{http://www.eso.org/instruments/naco/inst/filters.html}.

We obtained 20 images in the $J$ band, and 40 images in the $K$ band, each with an exposure time of 0.35\,s. This is the minimum full-frame integration time of CONICA. The $J$ band images were collected during $\approx 2$\,minutes around UT01:53:31, and the $K$ images in $\approx 4$\,minutes around UT01:58:20. During these observations the DIMM seeing at Paranal in the visible was good ($0.6-0.7\arcsec$), resulting in a high Strehl ratio {($\approx 45-50\%$)}.
The raw images were dark-subtracted, flat-fielded (using lamp flats), and corrected for bad pixels using IRAF\footnote{IRAF is distributed by the NOAO, which is operated by the Association of Universities for Research in Astronomy, Inc., under cooperative agreement with the National Science Foundation.}.
On the NACO images, we measured both the differential photometry and the differential astrometry of \betacrb~B relatively to \betacrb~A taken as the reference.

To measure the relative astrometry, we treated each image separately using 
the Yorick\footnote{http://yorick.sourceforge.net/} software package. 
We used a classical $\chi^2$ minimization to fit an extracted sub-image 
of \betacrb~A (with a size of $9 \times 9$\,pixels) at the position of 
the fainter component B. The interpolation of the shifted image of A was 
done in Fourier space. The adjusted parameters were the relative positions 
$d({\rm RA})$ and $d({\rm Dec})$, the flux ratio and the background level, 
although we used only the relative separations for our astrometric analysis. 
In order to estimate the associated error bars, we used the bootstrapping 
technique described by \cite{kervella04a}. This technique is also 
called ``sampling with replacement'' and consists of constructing a hypothetical, 
large population derived from the original measurements and estimate the statistical 
properties from this population. The technique allows us to compute meaningful 
confidence intervals without any assumption on the properties 
of the underlying population (\eg\ a Gaussian distribution).
We validated the adopted Fourier interpolation method by comparing the results with a simple 
Gaussian fit of the two PSF cores. The two methods yield exactly the same relative positions 
(within $150\,\mu$as), although the Gaussian fit has a slightly larger dispersion 
due to the mismatch of the slightly seeing-distorted PSF and the Gaussian function. 
We obtained the following vector separations along the RA and Dec directions 
of B relatively to A, for the epoch of the observations (MJD 54633.08):
\begin{equation}
d({\rm RA}) =   177.84 \pm 0.09 \pm 0.40 \,{\rm mas},
\end{equation}
\begin{equation}
d({\rm Dec}) =   -208.38 \pm 0.06 \pm 0.47\,{\rm mas}.
\end{equation}
The two stated error bars are the statistical and systematic uncertainties, respectively. The latter includes the pixel scale uncertainty and the detector orientation uncertainty. These values correspond to a separation $r$ and position angle $\alpha$ (east of north) of:
\begin{equation}
r = 273.95 \pm 0.63\,{\rm mas}, \label{eq:sep}
\end{equation}
\begin{equation}
\alpha = 139.52 \pm 0.13\,{\rm degrees}.
\end{equation}
These measurements were done on the $K$ images since they have the best Strehl ratio. In the $J$ band, 
the Strehl ratio was lower and more unstable, resulting in a significantly variable background from A to B. 
Although its average value is not a concern, its slope tends to slightly shift the average apparent position 
of $B$, by $-0.2$ and $-1.4$\,mas in RA and Dec, respectively (towards the lower left quadrant of Fig.~\ref{betcrbK}).  
The average NACO image in the $K$ band is presented in Fig.~\ref{betcrbK}, 
together with the orbit by \cite{tokovinin84}. Our astrometric measurement falls on the predicted orbit 
within only 7\,mas. Note that the reference epoch of the orbital elements by \cite{north98} appears 
to be late by approximately 300\,days.

%
\begin{table}
\caption{Interferometric calibrators selected from \cite{merand05}.}
\label{cals_table}
\centering
\begin{tabular}{cccc}
\hline \hline
       & Spectral &  $K$   &  UD diameter  \\
HD     & type     &  [mag] &  in K band [mas] \\
\hline
147266 & G8III    & $3.8$  & $0.785 \pm 0.011$\\ 
145457 & K0III    & $4.1$  & $0.693 \pm 0.009$\\ 
108123 & K0III    & $3.7$  & $0.929 \pm 0.012$\\ 
\hline
\end{tabular}
\end{table}
%

The photometry was obtained in two steps: (1) we obtained the combined photometry of the two stars; (2) we then measured the differential flux of B relative to A. We will discuss these steps in the following.

(1) The combined ADU count was computed from the NACO images by using a large aperture enclosing the full PSFs of the two stars. It was then converted to magnitude using the Zero Points obtained routinely by the observatory on the same night: ${\rm ZP}\,(J) = 24.062 \pm 0.048$ and ${\rm ZP}\,(K) = 22.985 \pm 0.078$, and an attenuation of $4.81 \pm 0.03$\,mag for the neutral density filter. These zero points are not corrected for atmospheric absorption, but as they were obtained at low airmass ($\approx 1.15$), we neglect the atmospheric absorption of $\approx 0.01$\,mag in $J$ and $K$. We corrected the atmospheric absorption using the standard values by \cite{nikolaev00}, namely 0.092\,mag/AM (relative to unit airmass) for $J$ and 0.065\,mag/AM for $K$, for our observation airmass of 1.71. We obtain:
\begin{equation}
m_J\,({\rm A}+{\rm B}) = 3.28 \pm 0.07,
\end{equation}
\begin{equation}
m_K\,({\rm A}+{\rm B}) = 3.22 \pm 0.08.
\end{equation}

(2) The differential photometry was obtained slightly differently, since the diffuse background of \betacrb~A tends to contaminate the flux of star B (but the reverse effect is negligible). We first computed aperture photometry of A on the average $J$ and $K$ images using very small aperture radii of 3\,pixels in the $J$ band and 4\,pixels in the $K$ band (contamination is lower in $K$). 
We calculated the median background value in concentric rings centered on A. This value was then subtracted from component B's flux.
This allowed us to subtract the diffuse light from the PSF wings of A at the position of B. We checked that the residual background around B was negligible. We then integrated the flux of B on the ring-median-subtracted image using the same aperture radius as for A. We obtain the following flux ratios of each star relative to the total of the two, \ie\ $\rho(\star) = f(\star) / f({\rm A}+{\rm B})$:
\begin{equation}
\rho_J\,({\rm A}) = 0.7957 \pm 0.0084,\ \ \rho_J\,({\rm B}) = 0.2043 \pm 0.0021,
\end{equation}
\begin{equation}
\rho_K\,({\rm A}) = 0.7785 \pm 0.0002,\ \rho_K\,({\rm B}) = 0.2215 \pm 0.0002.
\end{equation}
Note that the quoted uncertainties are statistical errors and do not include possible flat-fielding errors. 
From the combined magnitudes determined above, we obtain the individual magnitudes of \betacrb~A and B:
\begin{equation}
m_J\,({\rm A}) = 3.54 \pm 0.07,\ \ m_J\,({\rm B}) = 5.00 \pm 0.07, 
\end{equation}
\begin{equation}
m_K\,({\rm A}) = 3.50 \pm 0.08,\ m_K\,({\rm B}) = 4.86 \pm 0.08.
\end{equation}
The individual $J,K$ magnitudes have large uncertainties,
but we stress that we only use the values of $\rho$ for the interpretation of 
our interferometric data (Sect.~\ref{sec:chara}), 
and they are known with a much higher accuracy.

\subsection{CHARA/FLUOR interferometry\label{sec:chara}}

Our interferometric observations of \betacrb\ took place on 17--18 May 2008 
in the near infrared $K'$ band ($1.9\leq \lambda \leq  2.3\,\mu\mathrm{m}$) 
at the CHARA Array \citep{tenbrummelaar05} using FLUOR 
(the Fiber Linked Unit for Optical Recombination; \citealt{coude03}). 
We used the FLUOR Data Reduction Software 
(DRS; \citealt{coude97, kervella04a, merand06})  
to extract the squared instrumental visibility of the interference fringes. 
We used three different interferometric calibrators in order to calibrate 
the visibilities on sky. Their properties are listed in Table~\ref{cals_table}.
We note that the angular diameters of the calibrator stars 
are comparable to or larger than the target star. 
Therefore they contribute significantly to the uncertainty of the angular diameter measurement.
The corresponding systematic uncertainties were propagated into the final angular diameter uncertainties.

%
\begin{table}
\caption{Journal of observations.}
\label{obs_V2} \centering
\begin{tabular}{lccr r@{}l@{}l r@{}r}
\hline \hline 
& MJD & $B$  & $PA$ & \multicolumn{3}{c}{$V^2$} \%& \multicolumn{2}{c}{$V^2(\star)$ \% }   \\
& $-54000$ & (m) & (deg) & \multicolumn{3}{c}{observed} & \multicolumn{2}{c}{corrected} \\
\hline
A& $604.388$ & $324.09$  & $  6.17$ & $ 33.$&$2  $&$\pm1.6$ &$ 54.7\pm $&$2.7$\\ 
A& $604.408$ & $323.61$  & $  1.40$ & $ 31.$&$5  $&$\pm1.4$ &$ 52.0\pm $&$2.3$\\ 
A& $604.433$ & $323.88$  & $ -4.70$ & $ 29.$&$0  $&$\pm1.2$ &$ 47.8\pm $&$2.0$\\ 
A& $604.488$ & $327.21$  & $-17.40$ & $ 34.$&$1  $&$\pm3.6$ &$ 56.3\pm $&$5.9$\\ 
A& $605.296$ & $245.70$  & $-39.57$ & $ 45.$&$9  $&$\pm3.1$ &$ 75.8\pm $&$5.1$\\ 
A& $605.326$ & $248.92$  & $-45.62$ & $ 44.$&$7  $&$\pm2.2$ &$ 73.8\pm $&$3.7$\\ 
\hline
B& $604.398$ & $323.76$  & $  3.65$ & $  3.$&$44 $&$\pm0.25$ &$ 70.0\pm$&$ 5.2$\\
B& $605.254$ & $238.03$  & $-29.58$ & $  4.$&$76 $&$\pm1.38$ &$ 97.1\pm$&$ 28.0$\\
\hline
\multicolumn{9}{l}{\emph{Notes:} $B$ is the projected length of the interferometric baseline,}\\
\multicolumn{9}{l}{$PA$ is the projection angle of the baseline, $V^2$ is the observed}\\
\multicolumn{9}{l}{squared visibility and $V^2(\star)$ is the squared visibility corrected }\\
\multicolumn{9}{l}{for the presence of the other component (see Eq.~\ref{correction_factor}).}

\end{tabular}
\end{table}
%

The light from both stars of \betacrb\ is injected simultaneously in the FLUOR fibers, 
since the acceptance angle is $0.8\arcsec$ on the sky. However, due to their on-sky separation of $\approx 0.3\arcsec$ (\cf\ Eq.~\ref{eq:sep}), 
two fringe packets are formed at different optical path differences.

For this reason, we have to correct our measured visibility for this effect. The monochromatic visibility of the binary is:
\begin{equation}
V = \rho_K({\rm A}) \, V({\rm A}) + 
  \rho_K({\rm B}) \, V({\rm B}) \, e^{\,2\,i\,\pi\, \mathbf{B} \cdot \mathbf{\Gamma}\,/\, \lambda},
\end{equation}
where $\rho_K({\rm A})$ and $\rho_K({\rm B})$ are the relative fluxes of A and B,
$V({\rm A})$ and $V({\rm B})$ are the individual visibilities, $\mathbf{B}$ the
baseline vector, $\mathbf{\Gamma}$ the angular separation between A and B, and
$\lambda$ is the wavelength.  Because we observed
over a relatively broad wavelength range with FLUOR, 
and since the binary is well-resolved by our baselines,
the fringes appeared as two distinct fringe packets. Moreover, FLUOR has
a limited window in terms of optical path difference, corresponding to
a limited field of view. For $\beta$~CrB\ A$+$B, 
the two fringe packets are not present in a single fringe scan. 
Hence, the squared visibility measured by FLUOR in the case of A is
\begin{equation}
V^2 = \rho_K({\rm A})^2\,V^2({\rm A}).
\label{correction_factor}
\end{equation}
In the case of the observations of B, 
this multiplicative factor $\rho_K({\rm B})^2$ is 
small (due to the faintness of the star)
and causes an amplification of the error bars on the true visibility 
$V^2({\rm B})$ for our second observation of this star.
After reducing and calibrating the data with the DRS pipeline
for each component, we have corrected the visibilities for
this effect and then derived the angular diameters using
limb-darkened models from \cite{claret00}.

This leads to the following angular diameters: 
\begin{equation}
\theta_\mathrm{LD}({\rm A}) = 0.699\pm0.017\,\mathrm{mas}\ (2.4\percent),
\end{equation}
\begin{equation}
\theta_\mathrm{LD}({\rm B}) = 0.515\pm0.054\,\mathrm{mas}\ (10.5\percent).
\end{equation}
The angular diameter of \betacrb~B is significantly more uncertain than that of A,
due to the very low apparent visibility (``$V^2$ observed" in Table~\ref{obs_V2})
of its interference fringes ``on top" of the incoherent flux from A. The large multiplicative
factor $\rho_K({\rm B}) \approx 20$ (Eq.~\ref{correction_factor}) also applies to the associated
$V^2$ error bar. Only two measurements of B could be derived from our data, and one
of these points (MJD 54604.398) dominates the fit.
In the case of A, our six data points are in fair agreement with the best-fit model with a reduced $\chi^2$ of $2.0$.

%

\subsection{Surface brightness-colour predictions \label{sbc}}

We can compare the measured angular diameters of \betacrb~A and B with the predictions from the surface brightness-colour (hereafter SBC) relations calibrated by \cite{kervella04b} using their $(V, V-K)$ relation. The $K$ band magnitudes were obtained in Sect.~\ref{naco_obs}. We derive the $V$ band magnitudes from the total magnitude of the system of $m_V({\rm A}+{\rm B}) = 3.67\pm0.05$ \citep{rufener88}, and the magnitude difference $\Delta m = 1.99\pm0.10$ measured by \cite{horch04} by speckle interferometry at 503\,nm. We have adopted estimated uncertainties on $m_V$ and $\Delta m$ since they are not given explicitly in the references. This gives the component magnitudes in $V$:
\begin{equation}
m_V({\rm A}) = 3.83 \pm 0.05,\ \ m_V({\rm B}) = 5.82 \pm 0.10, \label{eq:mv}
\end{equation}
and the predicted photospheric angular diameters:
\begin{equation}
\theta_{\rm LD}({\rm A}) = 0.696 \pm 0.028\,{\rm mas}\ (4.0\percent),
\end{equation}
\begin{equation}
\theta_{\rm LD}({\rm B}) = 0.415 \pm 0.017\,{\rm mas}\ (4.1\percent).
\end{equation}
The predicted and measured angular diameters of \betacrb~A are therefore in excellent agreement, while they are in satisfactory agreement for B (difference of 1.7\,$\sigma$). We note than increasing the uncertainty on $m_V$ by a factor two does not significantly change the uncertainty of the angular diameters.


\begin{figure*}[t]
\centering
\includegraphics[width=17cm]{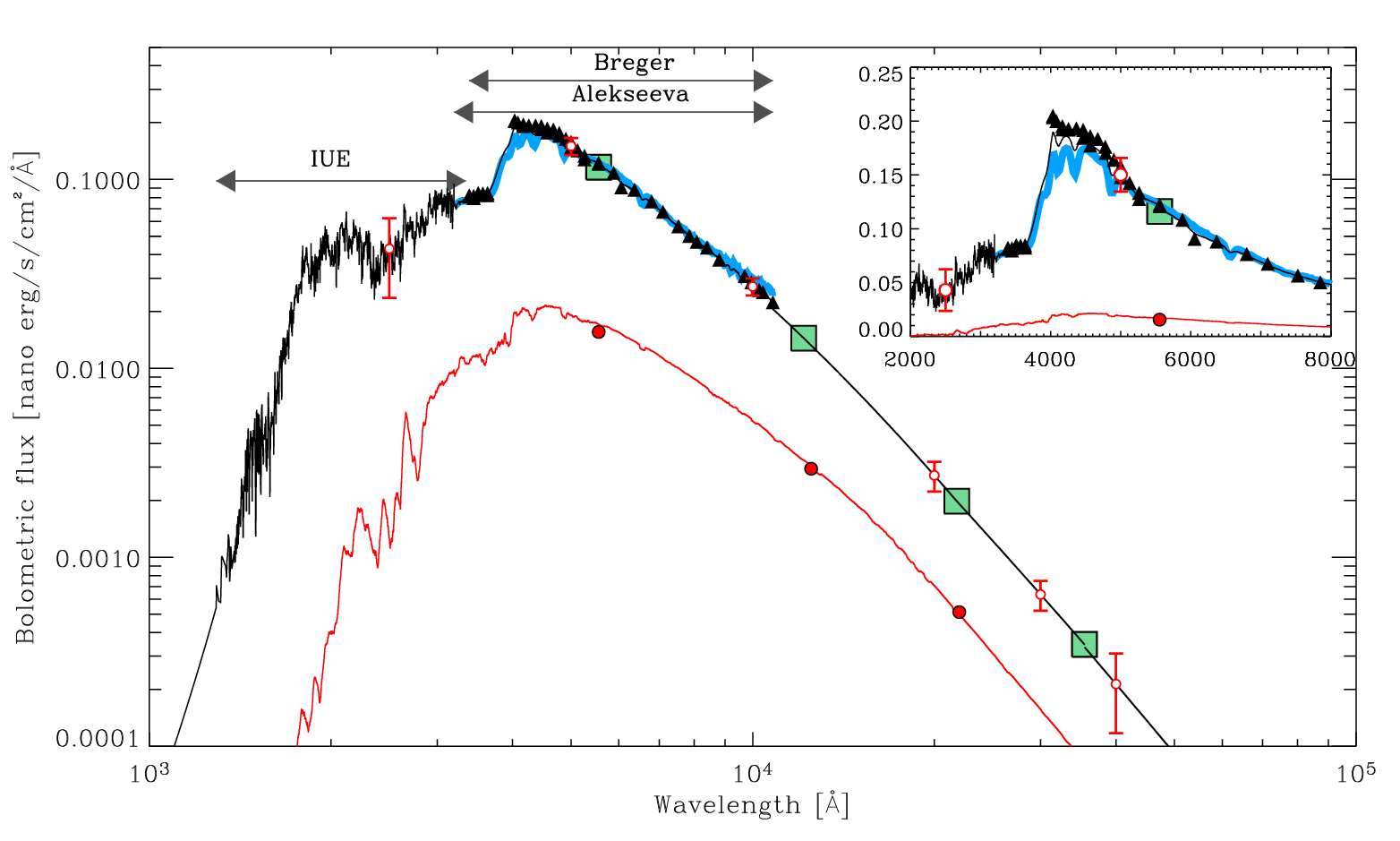} 
\caption{The solid black line shows the flux distribution for \betacrb~A$+$B, 
obtained by combining IUE satellite UV data, 
spectrophotometry from \cite{breger76} (triangles) and \cite{alekseeva96} (thick curve), 
and broad band fluxes from \cite{morel78} (box symbols).
The curve peaking at $\simeq0.02\times10^{-9}$\,erg/s/cm$^2$/\AA\ is 
the ATLAS9 model spectrum fitted to the B component, 
scaled by fitting the $V$ and NACO $J$,$K$ magnitudes (filled circles).
The open circles indicate the assumed 3-sigma error bars 
on the combined flux of the components at six different wavelengths.
The inset shows the details from 2000--8000~\AA\ on a linear flux scale.
\label{fig:spec}}
\end{figure*}

\subsection{Linear photospheric radii \label{sec:linrad}}

The original \emph{Hipparcos} parallax of \betacrb\ is $\pi = 28.60\pm0.69$\,mas \citep{esa97}, consistent with the new reduction by \cite{vanleeuwen07} of $\pi = 29.17 \pm 0.76$\,mas. However, as the new reduction is not corrected for binarity effects, 
we adopt the original \emph{Hipparcos} parallax.

For \betacrb~A, the angular diameter measurement presented in Sect.~\ref{sec:chara} represents a significant improvement in accuracy, by a factor of $1.6$, over the surface brightness-colour estimate of Sect.~\ref{sbc}. For the B component, this is not the case, as the visibility measurement is made particularly difficult by the presence of the brighter A component. For the subsequent analysis presented in Sect.~\ref{sec:analysis}, we therefore choose to adopt for \betacrb~A our direct interferometric angular diameter measurement, while for B we will use the SBC estimate computed from our $K$ band NACO photometry. This gives the following linear radii:
\begin{equation}
R_{\rm interf}({\rm A}) = 2.63 \pm 0.09\ {\rm R}_\odot\ (3.4\percent),
\end{equation}
\begin{equation}
R_{\rm SBC}({\rm B}) = 1.56 \pm 0.07\ {\rm R}_\odot\ (4.5\percent).
\end{equation}

\section{The effective temperatures and masses of \betacrbbold \label{sec:analysis}}

In the following we will determine the effective temperatures and luminosities
of the components of \betacrb\ using two methods. 
The first method (Sect.~\ref{sec:lumbc}) relies on the bolometric correction (model-dependent), 
while the second method (Sect.~\ref{sec:lumn}) is only weakly model-dependent.
We will then compare the radius and \teff\ of the components with a grid of isochrones
to determine their approximate age and evolutionary masses (Sect.~\ref{sec:age}).



\subsection{Luminosity and effective temperature from angular diameter $+$ BC $+$ parallax
\label{sec:lumbc}}

We used the bolometric corrections ($BC_V$) from \cite{bessell98}.
For the measured $V-K$ values of $0.36 \pm 0.09$ and $0.99 \pm 0.09$, for A and B
we get BC$_V({\rm A}) = 0.04$, BC$_V({\rm B}) = 0.01$, BC$_K({\rm A}) = 0.39$, and BC$_K({\rm B}) = 1.01$
(with $\log g = 4$). We assume the uncertainties on the BCs are 0.02 mag.
We therefore obtain from the $V$ band photometry:
\begin{equation}
m_{\rm bol}({\rm A}) = 3.87 \pm 0.05,\ \ \ m_{\rm bol}({\rm B}) = 5.83 \pm 0.10.
\end{equation}
The same computation with the $K$ band magnitudes gives identical values within 0.01\,mag.
We thus obtain the following bolometric luminosities, assuming the \emph{Hipparcos}
parallax and $M_{\rm bol}=4.75$ for the Sun (recommendation by IAU 1999):
\begin{equation}
L^{\rm BC}\,({\rm A}) = 27.6 \pm 1.8\ {\rm L}_\odot,\ \ \ L^{\rm BC}\,({\rm B}) = 4.5 \pm 0.5\ {\rm L}_\odot.
\end{equation}

We can now use the radii determined in Sect.~\ref{sec:linrad} to derive the
effective temperatures of the two stars through 
$L/{\rm L}_\odot = (R/{\rm R}_\odot)^2 \, (T_{\rm eff} / {\rm T}_{{\rm eff;}\odot})^4$, 
where we use the solar value ${\rm T}_{\rm eff;\odot}=5777$\,K \citep{cox2000}:
\begin{equation}
T_{\rm eff}^{\rm BC}\,({\rm A}) = 8160 \pm 200\ {\rm K},\ \ \ T_{\rm eff}^{\rm BC}\,({\rm B}) = 6750 \pm 230\ {\rm K}.
\end{equation}

\subsection{Effective temperature from angular diameter $+$ flux\label{sec:lumn}}


The above method for the determination of \teff\ has the caveat that it relies on 
the bolometric correction being valid for these stars. The BCs from \cite{bessell98}
are calculated from atmospheric models and do not depend on the metallicity. 
As a check, we will determine the \teff\ of the A component by a direct method,
meaning it will only be weakly dependent on the assumed model atmosphere. This is
done by calculating the integrated bolometric flux of ${\rm A}+{\rm B}$ and subtracting the
flux from the $B$ component using an ATLAS9 model with the \teff\ determined above.

The bolometric flux of the combined system, \betacrb\ A$+$B, was obtained by combining 
data from the literature from the UV to the near IR as shown in Fig.~\ref{fig:spec}.
In the UV range we used five spectrograms from the Sky Survey Telescope \citep{jamar76}
obtained at the \textit{IUE} ``Newly Extracted Spectra'' 
data archive\footnote{http://sdc.laeff.inta.es/cgi-ines/IUEdbsMY/}, covering the
wavelength interval 1150~\AA~$< \rm \lambda <$ 3350~\AA. 
We computed the weighted average of the spectrograms and some spurious data from 1150--1250~\AA\ were removed.
In the optical range we used spectrophotometry
from \cite{alekseeva96} and \cite{breger76}, which cover the range 
3200~\AA $< \rm \lambda <$ 10800~\AA.
In addition, for the near-IR wavelengths we used the broadband magnitudes $VJKL$
from \cite{morel78}. We interpolated the points between the broadband data 
and made a linear extrapolation at the end points (UV and IR ranges), 
although in practice the contribution is negligible. 
Finally, we calculated the weighted average flux, which is
shown as the solid black line in Fig.~\ref{fig:spec}. For the relative flux uncertainties we assumed 
15\percent\ in the UV, 6\percent\ in the optical, and 10\percent\ in the near IR. 
For the ranges where extrapolations were made, we doubled these errors.
The adopted uncertainties are larger than the originally published values.
We adjusted them based
on the disagreement between different sources of data in the same 
wavelength ranges, \ie\ the spectrophotometric data
from \cite{breger76} and \cite{alekseeva96} in Fig.~\ref{fig:spec}.




Since \betacrb\ is a binary system, extra care must be taken
when computing the bolometric flux of the primary star.
The binary has a maximum angular separation of {0.3''} and 
all available flux data contain the combined light of the two components.
Since our main interest is the A component, we have to estimate and 
subtract the flux of the B component.
To accomplish this, we fitted Kurucz
models to the $m_J$, $m_K$ and $m_V$ magnitudes of the $B$ component,
taking into account the statistical errors on the magnitudes 
(solid circles in Fig.~\ref{fig:spec}). For the $m_J$ and $m_K$ magnitudes 
we adopted the measurements from NACO, while the $m_V$ magnitude used 
was that derived in Eq.~\ref{eq:mv}. The spectra for the Kurucz models
were obtained using the IDL routine {\tt kurget1} (ATLAS9 models) and the corresponding
database of models available in the IUE reduction and data analysis package 
IUEDAC\footnote{http://archive.stsci.edu/iue/iuedac.html}.
We started by converting the $m_V$ magnitude of \betacrb~B into flux using the
relation (6) of \cite{rufe88} and used this result to
calibrate the models. We then converted the Kurucz fluxes at the NACO $J$ and 
$K$ central wavelengths (12,650~\AA\ and 21,800~\AA) into magnitudes using $m = -2.5\times \log_{10}(f/f_0)$.
Here $m$ is the magnitude in a given filter, $f$ is the flux at the central wavelength of
that filter and $f_0$ is the standard zeroth-magnitude flux for the same filter. The values
of $f_0$ were computed by integrating the flux of Vega through each of the $J$ and $K$ filters
of the NACO instrument, and assuming that Vega has zero magnitude 
in all bands\footnote{We note that \cite{bohlin04} recently found $V=0.026\pm0.008$ for Vega.}.
{We then generated 100 values for $m_J$ and $m_K$ of \betacrb~B 
by adding random fluctuations consistent with the uncertainties. 
For each set we determined the Kurucz model that best fitted 
each set of magnitudes. The average integrated flux of 
the 100 fitted Kurucz models was found to be $f_B=(1.2\pm0.3) \, 10^{-7}$ erg/s/cm$^2$. 
We adopted a rather large uncertainty (25\percent) since we only have three
broadband flux measurements of the B component.

The observed flux from the combined system was computed by integrating the
black curve shown in Fig.~\ref{fig:spec}, from which 
we obtained $f_A + f_B = (7.8\pm0.4)\,10^{-7}$ erg/s/cm$^2$
and the bolometric flux of the primary component is thus found to be
$f_A = 6.6\pm0.5\,10^{-7}$ erg/s/cm$^2$. We then obtained the effective
temperature using the relation,
\begin{equation}
\label{eq:theta}
\sigma\,T_{\rm eff}^4 = \frac{ 4\,f_{\rm bol}}{\theta_{\rm LD}^2},
\end{equation}
where $f_{\rm bol}$ is the bolometric flux and $\theta_{\rm LD}$ is the limb-darkened angular diameter. 
Inserting the values for $f_A$ and the angular diameter from Sec.~\ref{sec:chara}
we obtain an effective temperature of
\begin{equation}
T_{\rm eff}^\theta\,({\rm A}) = 7980\pm180\,{\rm K}.   
\end{equation}
Combining this with the radius we get the luminosity
\begin{equation}
L^\theta \, ({\rm A}) = 25.3\pm2.9\, {\rm L}_\odot.
\end{equation}
These values are in agreement with those in Sect.~\ref{sec:lumbc} where we
used the (model dependent) bolometric correction. 
Since the calculation using Eq.~\ref{eq:theta} is
nearly model-independent (limb-darkening coefficients depend on atmosphere models),
we adopt these values as our final estimates of \teff\ and the luminosity.
We note that if we neglect 
the contribution to the flux from the B component, 
$T_{\rm eff}^\theta\,({\rm A})$ becomes 350\,K higher.



Several determinations of \teff\ are found in the literature and we mention a few here.
\cite{koch06} used photometric indices to determine \teff\ and found $7430\pm200$\,K,
which is significantly lower than our value. 
\cite{netopil08} have determined the \teff\ of \betacrb\ from three photometric systems
(Str\"omgren, Geneva and Johnson) and compared these with values in the literature. 
The mean value for the photometric indices is $7710\pm260$\,K and the mean 
of the literature values is $8340\pm360$\,K. This is a typical example of the large scatter
found for chemically peculiar A stars.
However, since the rms scatter is large, the results summarized by \cite{netopil08} 
are in acceptable agreement with our new determination. 
It is worth stressing that our determination is the first that is not affected by
photometric calibration errors or interstellar reddening, 
and is only weakly dependent on the adopted limb darkening.


%
%
\begin{table}
\caption{Measured quantities and derived fundamental parameters.  
}
\label{tab:fund}
\centering
\begin{tabular}{lr@{}lr@{}l}
\hline \hline

Parameter                  & \multicolumn{2}{c}{A} & \multicolumn{2}{c}{B}  \\ \hline
$\pi$ [mas]                & $28.60$&$\pm0.69$  & \multicolumn{2}{c}{Same~as~A} \\
\thetald\ [mas]            & $0.699$&$\pm0.017$ & $0.415$&$\pm0.017$  \\
\fbol\ [$10^{-7}\,$erg/s/cm$^{2}$] & 
                             $  6.6$&$\pm0.5  $ & $ 1.2 $&$\pm0.3  $  \\ \hline
$R/{\rm R}_\odot$          & $ 2.63$&$\pm0.09 $ & $ 1.56$&$\pm0.07 $  \\
$L/{\rm L}_\odot$          & $ 25.3$&$\pm2.9  $ & $ 4.5 $&$\pm0.5  $  \\
 \teff\ [K]                & $ 7980$&$\pm180  $ & $ 6750$&$\pm230  $  \\ 
Spectral type & \multicolumn{2}{c}{A5} & \multicolumn{2}{c}{F2} \\ \hline
$M/{\rm M}_\odot$          & $2.09$&$\pm0.15$   & $1.40$&$\pm0.10$ \\
Age [Gyr]                  & $0.53$&$\pm0.10$   & \multicolumn{2}{c}{Same~as~A} \\ 

\hline
\end{tabular}
\end{table}
%

\subsection{The evolutionary status and mass of \betacrb
\label{sec:age}}

To investigate the evolutionary status of the two components of
\betacrb\ we have compared the observed \teff\ and radius 
with isochrones from the BASTI grid \citep{basti04} as
shown in Fig.~\ref{fig:hr}.
To transform the mass fraction of heavy elements ($Z$) 
of the isochrones to spectroscopic \feh\ values 
we used the solar value ${\rm Z}_\odot = 0.0156$ \citep{caffau09},
\ie\ ${\rm [Fe/H]} \simeq \log_{10} \, (Z_{\rm BASTI}/{\rm Z}_\odot)$.
We assumed an uncertainty of $\pm0.002$ on ${\rm Z}_\odot$, 
which corresponds to $\pm0.05$~dex on \feh.


In Fig.~\ref{fig:hr} we show two sets of isochrones with ${\rm [Fe/H]} = +0.10$ and $+0.28$. 
The higher metallicity appears to be in better agreement with the location of the B component. 
\cite{kurtz07} found [Fe-{\sc ii}/${\rm H]} = +0.41\pm0.22$ from 11 lines of singly ionized Fe.
This is the metallicity in the photosphere 
but we assume it represents the entire star. 
Since radial stratification of Fe is well-known to be present in roAp stars, 
our assumed metallicity is an approximation, 
but it seems to be supported by the agreement with the location of the stars 
in the radius--\teff\ diagram in Fig.~\ref{fig:hr}.
With this assumption we determine the age to be $0.53\pm0.10$~Gyr
and the masses of the components to be 
$M_{\rm A}/{\rm M}_\odot  = 2.09\pm0.15$ and 
$M_{\rm B}/{\rm M}_\odot  = 1.40\pm0.10$. 
These ``evolutionary masses'' are in good agreement 
with the dynamical masses determined by \cite{north98}:
$M_{\rm A}/{\rm M}_\odot  = 1.87\pm0.13$ and 
$M_{\rm B}/{\rm M}_\odot  = 1.356\pm0.073$.

%
%
\begin{figure}[t]
\hspace{-0.7cm}
\includegraphics[width=10cm, angle=0]{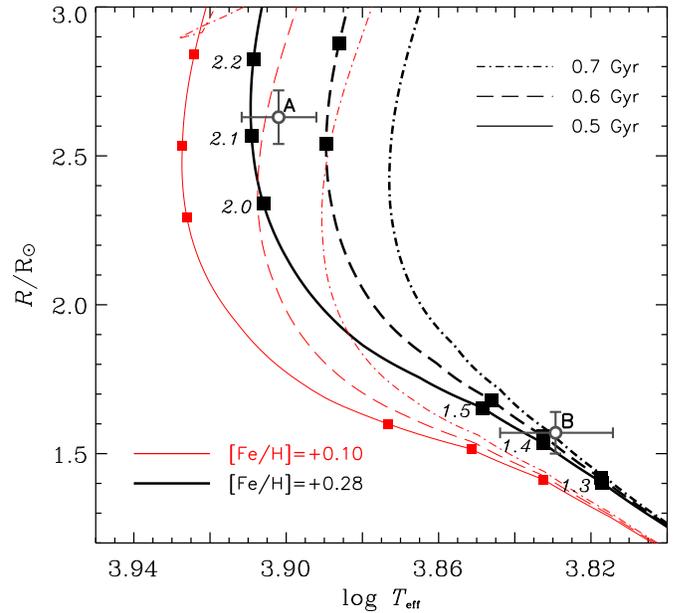} 
\caption{Radius--\teff\ diagram showing the location of the two components of \betacrb.
The thin lines are BASTI isochrones for ${\rm [Fe/H]} = +0.10$ for ages 0.5, 0.6 and 0.7 Gyr.
The thick lines are for the sames ages but for higher metallicity ${\rm [Fe/H]} = +0.28$.
On some of the isochrones the box symbols mark the masses at 
1.3--1.5\,\msun\ and 2.0--2.2\,\msun\ in steps of 0.1\,\msun.
\label{fig:hr}}
\end{figure}

\section{Discussion and conclusion \label{conclusion}}

We have determined the effective temperature of the 
primary component of the binary \betacrb\ using a technique that is
only weakly model-dependent. 
We used interferometric data to measure the angular diameter and the 
fluxes were constrained using NACO $J,K$ measurements of
each individual component in the binary.

We determined the primary component of \betacrb\ to have $T_{\rm eff}=7980\pm180$\,K.
In comparison, literature values for the combined star span $7230$--$8700$\,K 
(considering 1-$\sigma$ uncertainties).
From a similar analysis of flux data and interferometric data
the roAp star \acir, we found $T_{\rm eff}=7420\pm170$\,K \citep{bruntt08}.
For that star the literature values also span a large range
from $7470$--$8730$\,K. 
It is interesting that for \betacrb~A, our result is in the middle of the
range of previous estimates while for \acir\ the estimate is at the low end of the range.
If we compare solely with \teff\ estimates from the same photometric system,
\cite{koch06} found $7430\pm200$\,K for \betacrb\ and $7670\pm200$\,K for \acir.
We must remember that the photometric index of \betacrb\ includes both components,
and will always give a systematically low temperature. Taking this into account,
the photometric values from \cite{koch06} seem to agree with our fundamental 
(\ie\ model-independent) \teff\ values within about $\pm300$\,K.
It will be necessary to make interferometric measurements of several 
more of the brightest Ap stars to be able to improve the \teff\ scale 
of these peculiar stars.

Accurate determinations of \teff\ have important impact for the
asteroseismic modelling in future work.
\acir\ was observed for 84 days with the 52\,mm
star tracker on the now defunct WIRE satellite.
From the light curves \cite{bruntt09} detected five
frequencies of which two had not been observed before. 
These two lie symmetrically around the well-known dominant 
mode at 2442\,$\mu$Hz to form a triplet. 
\cite{bruntt09} interpreted the equidistant separation as half 
the large separation. Combining this with the new \teff,
the properties of the star could be constrained 
based on preliminary theoretical modelling of the observed pulsation modes.
To obtain similar results for \betacrb\ would be worthwhile
now that all ingredients for the modelling of the star are available, 
making it the second roAp star with well-established 
fundamental atmospheric parameters.
This would require an ambitious asteroseismic campaign \citep{kurtz07}
using a network of telescopes with stable spectrographs
like the Stellar Observations Network Group (SONG; \citealt{grundahl08}).

Our understanding of roAp stars would benefit from obtaining
interferometric angular diameters of more targets.  However, even with the
most sensitive beam combiners currently available, only a handful are
bright enough to yield a radius measurement to better than 2\percent.
Having now measured \acir\ and \betacrb, we next propose to observe \gammaequ\ and \tenaql.

\begin{acknowledgements}
The authors would like to thank all the CHARA Array and Mount Wilson Observatory 
daytime and nighttime staff for their support. 
The CHARA Array was constructed with funding from Georgia State University, the National 
Science Foundation, the W. M. Keck Foundation, and the David and Lucile Packard Foundation. 
The CHARA Array is operated by Georgia State University with support from the College of 
Arts and Sciences, from the Research Program Enhancement Fund administered by the Vice 
President for Research, and from the National Science Foundation under NSF Grant AST~0606958.
STR acknowledges partial support from NASA grant NNH09AK731.
MSC acknowledges the support of the Portuguese MCTES and of the FSE, 
of the European Union, through the programme POPH.
IMB would like to acknowledge the support from the Funda\c{c}\~ao para a Ci\^encia 
e~Tecnologia (Portugal) through the grant SFRH/BD/41213/2007.
LLK is supported by the Lend\"ulet program of the Hungarian Academy of Sciences.
This research has made use of the Washington Double Star Catalog maintained at the U.S. Naval Observatory.
This work received the support of PHASE, the high angular resolution partnership between
ONERA, Observatoire de Paris, CNRS and University Denis Diderot Paris 7.
This research took advantage of the SIMBAD and VIZIER databases at the CDS, Strasbourg 
(France), and NASA's Astrophysics Data System Bibliographic Services.
\end{acknowledgements}

\bibliographystyle{aa}
\bibliography{bruntt_betacrb} 

\begin{thebibliography}{49}
\expandafter\ifx\csname natexlab\endcsname\relax\def\natexlab#1{#1}\fi

\bibitem[{{Alekseeva} {et~al.}(1996){Alekseeva}, {Arkharov}, {Galkin},
  {Hagen-Thorn}, {Nikanorova}, {Novikov}, {Novopashenny}, {Pakhomov}, {Ruban},
  \& {Shchegolev}}]{alekseeva96}
{Alekseeva}, G.~A., {Arkharov}, A.~A., {Galkin}, V.~D., {et~al.} 1996, Baltic
  Astronomy, 5, 603

\bibitem[{{Bessell} {et~al.}(1998){Bessell}, {Castelli}, \& {Plez}}]{bessell98}
{Bessell}, M.~S., {Castelli}, F., \& {Plez}, B. 1998, \aap, 333, 231

\bibitem[{{Bohlin} \& {Gilliland}(2004)}]{bohlin04}
{Bohlin}, R.~C. \& {Gilliland}, R.~L. 2004, \aj, 127, 3508

\bibitem[{{Breger}(1976)}]{breger76}
{Breger}, M. 1976, \apjs, 32, 7

\bibitem[{{Bruntt} {et~al.}(2009){Bruntt}, {Kurtz}, {Cunha}, {Brand{\~a}o},
  {Handler}, {Bedding}, {Medupe}, {Buzasi}, {Mashigo}, {Zhang}, \& {van
  Wyk}}]{bruntt09}
{Bruntt}, H., {Kurtz}, D.~W., {Cunha}, M.~S., {et~al.} 2009, \mnras, 396, 1189

\bibitem[{{Bruntt} {et~al.}(2008){Bruntt}, {North}, {Cunha}, {Brand{\~a}o},
  {Elkin}, {Kurtz}, {Davis}, {Bedding}, {Jacob}, {Owens}, {Robertson}, {Tango},
  {Gameiro}, {Ireland}, \& {Tuthill}}]{bruntt08}
{Bruntt}, H., {North}, J.~R., {Cunha}, M., {et~al.} 2008, \mnras, 386, 2039

\bibitem[{{Caffau} {et~al.}(2009){Caffau}, {Maiorca}, {Bonifacio},
  {Faraggiana}, {Steffen}, {Ludwig}, {Kamp}, \& {Busso}}]{caffau09}
{Caffau}, E., {Maiorca}, E., {Bonifacio}, P., {et~al.} 2009, \aap, 498, 877

\bibitem[{{Campbell} \& {Moore}(1907)}]{moore1907}
{Campbell}, W.~W. \& {Moore}, J.~H. 1907, Lick Observatory Bulletin, 4, 162

\bibitem[{{Claret}(2000)}]{claret00}
{Claret}, A. 2000, \aap, 363, 1081

\bibitem[{{Coud\'e du Foresto} {et~al.}(2003){Coud\'e du Foresto}, {Borde},
  {M\'erand}, {Baudouin}, {Remond}, {Perrin}, {Ridgway}, {ten Brummelaar}, \&
  {McAlister}}]{coude03}
{Coud\'e du Foresto}, V., {Borde}, P.~J., {M\'erand}, A., {et~al.} 2003, in
  Society of Photo-Optical Instrumentation Engineers (SPIE) Conference Series,
  ed. W.~A. {Traub}, Vol. 4838, 280

\bibitem[{{Coud\'e Du Foresto} {et~al.}(1997){Coud\'e Du Foresto}, {Ridgway},
  \& {Mariotti}}]{coude97}
{Coud\'e Du Foresto}, V., {Ridgway}, S., \& {Mariotti}, J.-M. 1997, \aaps, 121,
  379

\bibitem[{{Cox}(2000)}]{cox2000}
{Cox}, A.~N. 2000, {Allen's astrophysical quantities} (Springer-Verlag)

\bibitem[{{Grundahl} {et~al.}(2008){Grundahl}, {Christensen-Dalsgaard},
  {Arentoft}, {Frandsen}, {Kjeldsen}, {J{\o}rgensen}, \&
  {Kjaergaard}}]{grundahl08}
{Grundahl}, F., {Christensen-Dalsgaard}, J., {Arentoft}, T., {et~al.} 2008,
  Communications in Asteroseismology, 157, 273

\bibitem[{{Hatzes} \& {Mkrtichian}(2004)}]{hatzes04}
{Hatzes}, A.~P. \& {Mkrtichian}, D.~E. 2004, \mnras, 351, 663

\bibitem[{{Horch} {et~al.}(2004){Horch}, {Meyer}, \& {van Altena}}]{horch04}
{Horch}, E.~P., {Meyer}, R.~D., \& {van Altena}, W.~F. 2004, \aj, 127, 1727

\bibitem[{{Jamar} {et~al.}(1976){Jamar}, {Macau-Hercot}, {Monfils}, {Thompson},
  {Houziaux}, \& {Wilson}}]{jamar76}
{Jamar}, C., {Macau-Hercot}, D., {Monfils}, A., {et~al.} 1976, {Ultraviolet
  bright-star spectrophotometric catalogue. A compilation of absolute
  spectrophotometric data obtained with the Sky Survey Telescope (S2/68) on the
  European Astronomical Satellite TD-1}

\bibitem[{{Kamper} {et~al.}(1990){Kamper}, {McAlister}, \&
  {Hartkopf}}]{kamper90}
{Kamper}, K.~W., {McAlister}, H.~A., \& {Hartkopf}, W.~I. 1990, \aj, 100, 239

\bibitem[{{Kervella} {et~al.}(2004{\natexlab{a}}){Kervella}, {S{\'e}gransan},
  \& {Coud{\'e} du Foresto}}]{kervella04a}
{Kervella}, P., {S{\'e}gransan}, D., \& {Coud{\'e} du Foresto}, V.
  2004{\natexlab{a}}, \aap, 425, 1161

\bibitem[{{Kervella} {et~al.}(2004{\natexlab{b}}){Kervella}, {Th{\'e}venin},
  {Di Folco}, \& {S{\'e}gransan}}]{kervella04b}
{Kervella}, P., {Th{\'e}venin}, F., {Di Folco}, E., \& {S{\'e}gransan}, D.
  2004{\natexlab{b}}, \aap, 426, 297

\bibitem[{{Kochukhov} \& {Bagnulo}(2006)}]{koch06}
{Kochukhov}, O. \& {Bagnulo}, S. 2006, \aap, 450, 763

\bibitem[{{Kochukhov} {et~al.}(2002){Kochukhov}, {Landstreet}, {Ryabchikova},
  {Weiss}, \& {Kupka}}]{koch02}
{Kochukhov}, O., {Landstreet}, J.~D., {Ryabchikova}, T., {Weiss}, W.~W., \&
  {Kupka}, F. 2002, \mnras, 337, L1

\bibitem[{{Kochukhov} {et~al.}(2008){Kochukhov}, {Ryabchikova}, {Bagnulo}, \&
  {Lo Curto}}]{koch08}
{Kochukhov}, O., {Ryabchikova}, T., {Bagnulo}, S., \& {Lo Curto}, G. 2008,
  Contributions of the Astronomical Observatory Skalnate Pleso, 38, 423

\bibitem[{{Kreidl}(1991)}]{kreidl91}
{Kreidl}, T.~J. 1991, \mnras, 248, 701

\bibitem[{{Kurtz}(1989)}]{kurtz89}
{Kurtz}, D.~W. 1989, \mnras, 238, 261

\bibitem[{{Kurtz} {et~al.}(2007){Kurtz}, {Elkin}, \& {Mathys}}]{kurtz07}
{Kurtz}, D.~W., {Elkin}, V.~G., \& {Mathys}, G. 2007, \mnras, 380, 741

\bibitem[{{Lenzen} {et~al.}(1998){Lenzen}, {Hofmann}, {Bizenberger}, \&
  {Tusche}}]{lenzen98}
{Lenzen}, R., {Hofmann}, R., {Bizenberger}, P., \& {Tusche}, A. 1998, in
  Society of Photo-Optical Instrumentation Engineers (SPIE) Conference Series,
  ed. A.~M. {Fowler}, Vol. 3354, 606

\bibitem[{{Masciadri} {et~al.}(2003){Masciadri}, {Brandner}, {Bouy}, {Lenzen},
  {Lagrange}, \& {Lacombe}}]{masciadri03}
{Masciadri}, E., {Brandner}, W., {Bouy}, H., {et~al.} 2003, \aap, 411, 157

\bibitem[{{M{\'e}rand} {et~al.}(2005){M{\'e}rand}, {Bord{\'e}}, \& {Coud{\'e}
  Du Foresto}}]{merand05}
{M{\'e}rand}, A., {Bord{\'e}}, P., \& {Coud{\'e} Du Foresto}, V. 2005, \aap,
  433, 1155

\bibitem[{{M{\'e}rand} {et~al.}(2006){M{\'e}rand}, {Coud{\'e} du Foresto},
  {Kellerer}, {ten Brummelaar}, {Reess}, \& {Ziegler}}]{merand06}
{M{\'e}rand}, A., {Coud{\'e} du Foresto}, V., {Kellerer}, A., {et~al.} 2006, in
  Society of Photo-Optical Instrumentation Engineers (SPIE) Conference Series,
  Vol. 6268

\bibitem[{{Morel} \& {Magnenat}(1978)}]{morel78}
{Morel}, M. \& {Magnenat}, P. 1978, \aaps, 34, 477

\bibitem[{{Muterspaugh} {et~al.}(2006){Muterspaugh}, {Lane}, {Kulkarni},
  {Burke}, {Colavita}, \& {Shao}}]{muterspaugh06}
{Muterspaugh}, M.~W., {Lane}, B.~F., {Kulkarni}, S.~R., {et~al.} 2006, \apj,
  653, 1469

\bibitem[{{Netopil} {et~al.}(2008){Netopil}, {Paunzen}, {Maitzen}, {North}, \&
  {Hubrig}}]{netopil08}
{Netopil}, M., {Paunzen}, E., {Maitzen}, H.~M., {North}, P., \& {Hubrig}, S.
  2008, \aap, 491, 545

\bibitem[{{Neubauer}(1944)}]{neubauer44}
{Neubauer}, F.~J. 1944, \apj, 99, 134

\bibitem[{{Nikolaev} {et~al.}(2000){Nikolaev}, {Weinberg}, {Skrutskie},
  {Cutri}, {Wheelock}, {Gizis}, \& {Howard}}]{nikolaev00}
{Nikolaev}, S., {Weinberg}, M.~D., {Skrutskie}, M.~F., {et~al.} 2000, \aj, 120,
  3340

\bibitem[{{North} {et~al.}(1998){North}, {Carquillat}, {Ginestet}, {Carrier},
  \& {Udry}}]{north98}
{North}, P., {Carquillat}, J.-M., {Ginestet}, N., {Carrier}, F., \& {Udry}, S.
  1998, \aaps, 130, 223

\bibitem[{{Oetken} \& {Orwert}(1984)}]{oetken84}
{Oetken}, L. \& {Orwert}, R. 1984, Astronomische Nachrichten, 305, 317

\bibitem[{{Perryman} \& {ESA}(1997)}]{esa97}
{Perryman}, M.~A.~C. \& {ESA}, eds. 1997, ESA Special Publication, Vol. 1200,
  {The HIPPARCOS and TYCHO catalogues. Astrometric and photometric star
  catalogues derived from the ESA HIPPARCOS Space Astrometry Mission}

\bibitem[{{Pietrinferni} {et~al.}(2004){Pietrinferni}, {Cassisi}, {Salaris}, \&
  {Castelli}}]{basti04}
{Pietrinferni}, A., {Cassisi}, S., {Salaris}, M., \& {Castelli}, F. 2004, \apj,
  612, 168

\bibitem[{{Renson} \& {Manfroid}(2009)}]{renson09}
{Renson}, P. \& {Manfroid}, J. 2009, \aap, 498, 961

\bibitem[{{Rousset} {et~al.}(2003){Rousset}, {Lacombe}, {Puget}, {Hubin},
  {Gendron}, {Fusco}, {Arsenault}, {Charton}, {Feautrier}, {Gigan}, {Kern},
  {Lagrange}, {Madec}, {Mouillet}, {Rabaud}, {Rabou}, {Stadler}, \&
  {Zins}}]{rousset03}
{Rousset}, G., {Lacombe}, F., {Puget}, P., {et~al.} 2003, in Society of
  Photo-Optical Instrumentation Engineers (SPIE) Conference Series, ed. P.~L.
  {Wizinowich} \& D.~{Bonaccini}, Vol. 4839, 140

\bibitem[{{Rufener}(1988)}]{rufener88}
{Rufener}, F. 1988, {Catalogue of stars measured in the Geneva Observatory
  photometric system : 4 : 1988} (Sauverny: Observatoire de Geneve, 1988)

\bibitem[{{Rufener} \& {Nicolet}(1988)}]{rufe88}
{Rufener}, F. \& {Nicolet}, B. 1988, \aap, 206, 357

\bibitem[{{Ryabchikova} {et~al.}(2004){Ryabchikova}, {Nesvacil}, {Weiss},
  {Kochukhov}, \& {St{\"u}tz}}]{ryab04}
{Ryabchikova}, T., {Nesvacil}, N., {Weiss}, W.~W., {Kochukhov}, O., \&
  {St{\"u}tz}, C. 2004, \aap, 423, 705

\bibitem[{{S{\"o}derhjelm}(1999)}]{soderhjelm99}
{S{\"o}derhjelm}, S. 1999, \aap, 341, 121

\bibitem[{{ten Brummelaar} {et~al.}(2005){ten Brummelaar}, {McAlister},
  {Ridgway}, {Bagnuolo}, {Turner}, {Sturmann}, {Sturmann}, {Berger}, {Ogden},
  {Cadman}, {Hartkopf}, {Hopper}, \& {Shure}}]{tenbrummelaar05}
{ten Brummelaar}, T.~A., {McAlister}, H.~A., {Ridgway}, S.~T., {et~al.} 2005,
  \apj, 628, 453

\bibitem[{{Tokovinin}(1984)}]{tokovinin84}
{Tokovinin}, A.~A. 1984, Pisma Astronomicheskii Zhurnal, 10, 293

\bibitem[{{Trilling} {et~al.}(2007){Trilling}, {Stansberry}, {Stapelfeldt},
  {Rieke}, {Su}, {Gray}, {Corbally}, {Bryden}, {Chen}, {Boden}, \&
  {Beichman}}]{trilling07}
{Trilling}, D.~E., {Stansberry}, J.~A., {Stapelfeldt}, K.~R., {et~al.} 2007,
  \apj, 658, 1289

\bibitem[{{van Leeuwen}(2007)}]{vanleeuwen07}
{van Leeuwen}, F., ed. 2007, Astrophysics and Space Science Library, Vol. 250,
  {Hipparcos, the New Reduction of the Raw Data}

\bibitem[{{Weiss} \& {Schneider}(1989)}]{weiss89}
{Weiss}, W.~W. \& {Schneider}, H. 1989, \aap, 224, 101

\end{thebibliography}

\end{document}